\documentclass[12pt]{article}

\usepackage[dvips]{graphicx}
\usepackage{epsfig}
\usepackage{amsmath,amsfonts,amssymb,amsthm}
\usepackage{mathrsfs,mathtools}
\usepackage{verbatim}
\usepackage{psfrag}
\usepackage{bm}
\usepackage{bbm}
\usepackage[utf8]{inputenc}
\usepackage[square,comma,sort&compress,numbers]{natbib}
\usepackage[dvipsnames]{xcolor}
\usepackage{slashed}
\usepackage{upgreek}
\usepackage{enumitem}
\usepackage{float}
\usepackage[T1]{fontenc}
\usepackage{soul}
\usepackage{pgfplots}
\usepackage{extarrows}
\pgfplotsset{compat=1.18}
\usetikzlibrary{external,decorations.pathmorphing}
\usepackage{cases}
\usepackage{empheq}
\usepackage{physics}
\usepackage{epsf,epsfig}
\usepackage{graphics}
\usepackage{subcaption}
\usepackage{lmodern}
\usepackage{tikz}
\usepackage{booktabs}
\usepackage[normalem]{ulem}
\usepackage[colorlinks=true,urlcolor=blue,anchorcolor=blue,citecolor=blue,filecolor=blue
,linkcolor=blue,menucolor=blue,linktocpage=true,pdfproducer=medialab,pdfa=true
]{hyperref}
\usepackage[T1]{fontenc}
\usepackage[errorstop]{feynmp}
\usepackage{lettrine}
\usepackage{Zallman}

\tikzset{
  external/only named=true,
  thick/.style={line width=.5pt},
  approximation/.style={line width=1.2pt},
  numerics/.style={black, dotted, line width=.8pt},
  amplitude/.style={dashed},
  estimate/.style={dashed, line width=.8pt},
  normal plot/.style={line width=.8pt},
}

\tikzset{snake it/.style={decorate, decoration=snake}}

\newcommand\blfootnote[1]{%
  \begingroup
  \renewcommand\thefootnote{}\footnote{#1}%
  \addtocounter{footnote}{-1}%
  \endgroup
}

\def\d{\mathrm{d}}

\def\A{\mathcal{A}}

\def\O{\mathcal{O}}

\newcommand{\rom}[1]{\uppercase\expandafter{\romannumeral #1\relax}}

\newcommand{\tv}{\mathrm{TV}}
\newcommand{\fv}{\mathrm{FV}}

\newcommand{\fo}{\mathrm{fo}}
\newcommand{\barrier}{\mathrm{bar}}

\def\bal#1\eal{\begin{align}#1\end{align}}

\begin{document}                                                                                                                      

\thispagestyle{empty}



\begin{center}
\Large\bf\boldmath
Dynamical criterion for biased domain-wall formation
\unboldmath
\end{center}

\vspace{-0.2cm}

\begin{center}
Wen-Yuan Ai$^{1,2,*}$\blfootnote{$^*$wenyuanai@sjtu.edu.cn} 
\vskip0.4cm

{\it $^1$State Key Laboratory of Dark Matter Physics,\\ Tsung-Dao Lee Institute and School of Physics and Astronomy,\\ Shanghai Jiao Tong University, Shanghai 201210, China\\}

{\it
$^2$Key Laboratory for Particle Astrophysics and Cosmology (MOE),\\ and Shanghai Key Laboratory for Particle Physics and Cosmology,\\ Shanghai Jiao Tong University, Shanghai 201210, China 
}
\vskip1.cm
\end{center}

\begin{abstract}
In the presence of a bias term, the conventional condition for forming a domain-wall network is $p_{\rm FV}>p_c\simeq 0.31$, with $p_{\rm FV}/(1-p_{\rm FV})=\exp(-\Delta V(0)/V_b(0))$, where $p_{\rm FV}$ is the false-vacuum fraction immediately after the phase transition, $\Delta V(0)$ is the zero-temperature energy splitting between the false and true vacua and $V_b(0)$ is the zero-temperature barrier height measured from the true vacuum. This criterion, however, cannot be generally valid, since it is insensitive to the dynamics of the phase transition. In this work, we derive a dynamical criterion for domain wall formation in the presence of a bias term. We evaluate $p_{\rm FV}$ at the freeze-out temperature of the false-vacuum correlation volumes $T_{\rm fo}$, obtaining a substantially stricter criterion. The same dynamical picture also yields a necessary consistency condition for applying scaling-regime gravitational-wave estimates, $T_{\rm fo}>T_{\rm ann}$, where $T_{\rm ann}$ is the annihilation temperature inferred from the scaling-regime dynamics.
\end{abstract}

\newpage
%
\hrule
\tableofcontents
\vskip.85cm
\hrule


\section{Introduction}
Domain walls are produced when a field theory with disconnected vacua undergoes a symmetry-breaking transition in the early Universe. They have far-reaching consequences and have long been a focus of cosmology~\cite{Vilenkin:1984ib,vilenkin1994cosmic}. For some recent work, see, e.g., Refs.~\cite{Liu:2019lul,Blasi:2022woz,Wei:2022poh,Wu:2022stu,Wu:2022tpe,An:2023idh,Blasi:2023sej,Sassi:2023cqp,Li:2023gil,Gouttenoire:2023gbn,Dunsky:2024zdo,Sassi:2024cyb,Lu:2024ngi,Ghoshal:2025gci,Gouttenoire:2025ofv,Li:2025gld,Lee:2025yvn,Kanemura:2025ixp,Ma:2025bjf,Blasi:2025tmn,Fu:2025qhf,Borah:2025bfa,Borah:2026kfo,Masubuchi:2026eau,Azzola:2026cwa,Roshan:2026yon,Guo:2026cuv,Vanvlasselaer:2026fay,Bai:2026udq,Atta:2026pvw,Datta:2026fav,Datta:2026ffs}.
In the absence of an efficient decay mechanism, their energy density redshifts more slowly than radiation and matter, leading to the cosmological domain-wall problem~\cite{Zeldovich:1974uw,Kibble:1976sj,Vilenkin:1981zs,Sikivie:1982qv,Vilenkin:1982ks,vilenkin1994cosmic}. A common solution is to make the symmetry approximate rather than exact, so that a small explicit breaking lifts the vacuum degeneracy. This idea has been explored extensively~\cite{Gelmini:1988sf,Coulson:1995nv,Larsson:1996sp,Hiramatsu:2012sc,Correia:2014kqa,Correia:2018tty,Krajewski:2021jje}. Long-lived domain-wall networks, especially those that decay after entering the scaling regime~\cite{Press:1989yh,Hindmarsh:1996xv,Garagounis:2002kt,Oliveira:2004he,Avelino:2005kn,Leite:2011sc}, can source sizable stochastic gravitational waves~\cite{Hiramatsu:2010yz,Hiramatsu:2013qaa,Saikawa:2017hiv}.

Most of this literature addresses the dynamics of a wall network after it has formed. In this approach, the bias enters as a pressure difference $\Delta V(0)$ across the wall, and the main question is whether the walls disappear before they dominate the Universe and before big-bang nucleosynthesis (BBN). This is the right question if the network is already known to exist. However, it leaves open a logically earlier question: does a biased transition actually produce domain walls in the first place?

This question was already recognized in the seminal work of Gelmini, Gleiser and Kolb~\cite{Gelmini:1988sf}. They used percolation arguments to require that both vacua be populated above the percolation threshold $p_c$. The ratio between the false-vacuum population and the true-vacuum population is often estimated as $p_{\fv}/p_{\tv}\sim \exp[-\Delta V(0)/V_b(0)]$, where $\Delta V(0)$ is the zero-temperature energy splitting between false and true vacua and $V_b(0)$ is the zero-temperature barrier height above the true vacuum. For a simple cubic lattice one has $p_c\simeq0.31$. Requiring $p_{\fv}>p_c$ then gives
\begin{align}
    \Delta V(0)/V_b(0)\lesssim \ln[(1-p_c)/p_c]\simeq 0.8\,.
\end{align}

The above criterion cannot be generally valid, since it contains no information about the dynamics of the phase transition. For example, if the temperature decreases infinitely slowly, the field should follow the true vacuum adiabatically, and no domain walls should form. The purpose of this paper is to propose a dynamical criterion for domain wall formation in the presence of a bias term. Although a related analysis has been given for condensed matter systems~\cite{Rams:2019njq,Rysti:2019fwr}, a systematic analysis for cosmological systems has been missing so far. Compared to string formation~\cite{Fujikura:2024biv}, domain wall formation in the presence of a bias is simple enough to allow a semi-analytic analysis.

The paper is organised as follows. In the next section, we discuss biased $Z_2$ symmetry breaking and review the conventional criterion for domain-wall formation. In Section~\ref{sec:new-criterion}, we derive a dynamical formation criterion that incorporates the evolution of the phase transition. In Section~\ref{sec:scaling-regime}, we discuss an implication of the theoretical development for the scaling-regime description of domain-wall gravitational-wave signals. We conclude in Section~\ref{sec:conclusions}.

\section{Conventional formation criterion}
\label{sec:model}

As an illustrative example, we consider biased $Z_2$ symmetry breaking, described by a real scalar field $\phi$ with the following finite-temperature effective potential
\begin{equation}
 V(\phi,T)=\frac{\lambda}{4}\left[\phi^2-v^2(T)\right]^2-\varepsilon\phi,
 \label{eq:z2_potential}
\end{equation}
up to a $\phi$-independent contribution,
where $\lambda, \varepsilon>0$. The positive tilt makes the positive-$\phi$ vacuum the true vacuum. For concreteness, we take $v^2(T)=v_0^2(1-T^2/T_c^2)$ below $T_c$, but most of the discussion only requires that $v(T)$ grows as the transition proceeds.

\begin{figure}
    \centering
    \includegraphics[width=0.65\linewidth]{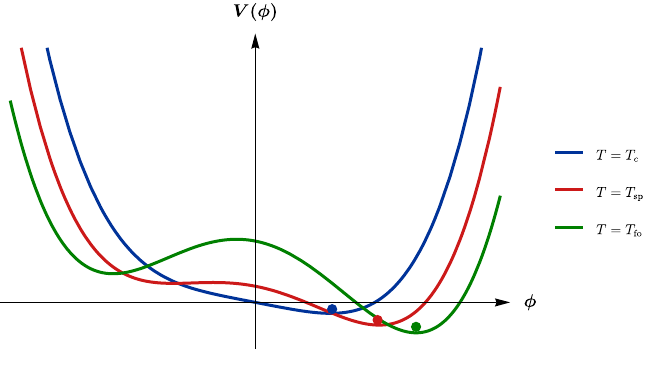}
    \caption{Illustration of the evolution of the potential below $T_c$. The dots track the true-vacuum minimum.
  The false vacuum appears only after the system crosses the spinodal temperature $T_{\rm sp}$, at which point false-vacuum correlation cells can be thermally populated. These cells eventually freeze out at $T_{\rm fo}$.
}
    \label{fig:pot}
\end{figure}

The potential has a unique minimum at sufficiently high temperature. A second local minimum, corresponding to the false vacuum, appears only after the system crosses a spinodal temperature $T_{\rm sp}<T_c$. See Fig.\,\ref{fig:pot} for an illustration. It is useful to introduce the dimensionless local tilt parameter
\begin{equation}
 \delta(T)\equiv \frac{\varepsilon}{\lambda v^3(T)}.
 \label{eq:delta_def}
\end{equation}
The existence of a false vacuum requires the cubic equation $\lambda[\phi^3-v^2(T)\phi]-\varepsilon=0$ to have three real roots. This gives
\begin{equation}
 \delta(T)<\delta_{\rm sp}=\frac{2}{3\sqrt3}.
 \label{eq:spinodal_condition}
\end{equation}
Thus, for fixed $\varepsilon$, the false vacuum appears only after $v(T)$ exceeds $v_{\rm sp}=(3\sqrt3\varepsilon/2\lambda)^{1/3}$. The discussion below applies after this spinodal point, when both local minima exist.

For $\delta(T)\ll 1$, the true and false minima are approximately $\phi_{\tv}(T)\approx v(T)$ and $\phi_{\fv}(T)\approx -v(T)$, while the barrier is located near $\phi_{\barrier}\approx 0$. The corresponding energy splitting and false-vacuum barrier height are $\Delta V(T)\equiv V_{\fv}(T)-V_{\tv}(T)\approx 2\varepsilon v(T)$ and $V_{b,\fv}(T)=V_{\rm bar}(T)-V_{\fv}(T)\approx \lambda v^4(T)/4-\varepsilon v(T)$. We denote the barrier height above the true vacuum as $V_b(T)$, i.e., without an additional subscript. Then $V_b(T)\approx \lambda v^4(T)/4+\varepsilon v(T)$.

Below $T_{\rm sp}$, false-vacuum cells of characteristic volume $\xi^3(T)$ can be thermally populated, where $\xi(T)$ is the correlation length evaluated in the true vacuum. In a simple mean-field treatment, one may estimate the correlation length as $\xi(T)\sim 1/m(T)$ with $m^2(T)=V''(\phi_{\tv},T)\simeq2\lambda v^2(T)$. The equilibrium ratio between false- and true-vacuum correlation volumes can be estimated as
\begin{align}
\label{eq:pfv-over-ptv}
\frac{p_{\fv}(T)}{p_{\tv}(T)}
\approx
\exp\left[
-\frac{\Delta V(T)\xi^3(T)}{T}
\right].
\end{align}
One might wonder whether a surface-energy term should be included in Eq.\,\eqref{eq:pfv-over-ptv}. Such a term would suppress false-vacuum cells even in the unbiased limit and would therefore fail to recover the consistency limit $p_{\fv}=p_{\tv}$ as $\Delta V\to 0$.

The conventional criterion for domain wall formation is obtained by evaluating Eq.\,\eqref{eq:pfv-over-ptv} at the Ginzburg temperature, which satisfies $V_b(T_G)\xi^3(T_G)/T_G\approx 1$, and replacing finite-temperature quantities by their zero-temperature values. Then we have
\begin{equation}
 \frac{p_{\fv}}{p_{\tv}}
 \approx \exp\left(-\frac{\Delta V(0)}{V_b(0)}\right).
 \label{eq:naive_prob}
\end{equation}
Requiring $p_{\fv}>p_c$ gives
\begin{equation}
 \frac{\Delta V(0)}{V_b(0)}\lesssim L_p\equiv \ln\left(\frac{1-p_c}{p_c}\right)\approx 0.8.
 \label{eq:naive_bound_general}
\end{equation}
For the quartic potential \eqref{eq:z2_potential}, this becomes approximately $8\varepsilon/(\lambda v_0^3)\lesssim 0.8$.

The above estimate assumes that the relevant false-vacuum population freezes when the barrier exponent is of order unity and that zero-temperature quantities can be used. Below, we show that neither assumption is generally correct in an expanding Universe.

\section{Dynamical freeze-out criterion}
\label{sec:new-criterion}

Once false-vacuum correlation cells populate, they need not be permanent.  If thermal activation over the barrier is rapid compared with the rate at which the potential changes, a false-vacuum domain will repeatedly convert into a true-vacuum domain, and the population will continue to track the instantaneous equilibrium value~\eqref{eq:pfv-over-ptv}. The population relevant for percolation should therefore be evaluated when these conversions freeze out.

The transition rate of a false-vacuum cell to a true-vacuum cell can be estimated as 
\begin{equation}
 \Gamma_{\fv\to\tv}(T)
 \simeq T
 \exp\left[-\frac{V_{b,\fv}(T)\xi^3(T)}{T}\right].
 \label{eq:Gamma_FT}
\end{equation}
This rate should not be confused with the thermal bubble-nucleation rate in a macroscopic false vacuum.  It is the local Ginzburg-cell transition rate that controls whether the occupation of a correlation cell continues to track equilibrium. The population freeze-out temperature is given by
\begin{equation}
 \Gamma_{\fv\to\tv}(T_{\rm fo})\simeq H(T_{\rm fo}).
 \label{eq:Tfo_def}
\end{equation}
Equivalently,
\begin{equation}
\label{eq:fo-cond}
 \boxed{\frac{V_{b,\fv}(T_{\rm fo})\xi^3(T_{\rm fo})}{T_{\rm fo}}
 \simeq\ln\left(\frac{T_{\rm fo}}{H(T_{\rm fo})}\right)\equiv L_\fo.}
\end{equation}
With the Friedmann equation
\begin{align}
\label{eq:Friedman}
    H=1.66\sqrt{g_{\star,\rho}}\frac{T^2}{M_{\rm Pl}},
\end{align}
where $M_{\rm Pl}\approx 1.22\times10^{19}\,{\rm GeV}$ is the Planck mass and $g_{\star,\rho}$ is the number of energy degrees of freedom, we have
\begin{align}
    L_\fo=36.5+\ln\left[\sqrt{\frac{100}{g_{\star,\rho}}}\left(\frac{100\,{\rm GeV}}{T_\fo}\right)\right].
\end{align}
Therefore, the freeze-out condition is much more stringent than the crude Ginzburg condition $V_{b}(T)\xi^3(T)/T=1$.

The false-vacuum fraction at freeze-out is
\begin{equation}
 \frac{p_{\fv}}{p_{\tv}}\bigg|_{T_{\rm fo}}
 =\exp\left(-\frac{\Delta V(T_{\rm fo})\xi^3(T_{\rm fo})}{T_{\rm fo}}\right).
 \label{eq:Afo_def}
\end{equation}
Demanding $p_{\fv}(T_{\rm fo})>p_c$ gives the refined formation criterion
\begin{equation}
 \boxed{
 \frac{\Delta V(T_{\rm fo})}{V_{b,\fv}(T_{\rm fo})}
 \lesssim \frac{L_p}{L_\fo},
 }
 \label{eq:refined_main}
\end{equation}
where $L_p$ is defined in Eq.~\eqref{eq:naive_bound_general}. This criterion is parametrically stronger than the naive zero-temperature estimate.

In the adiabatic limit $H\to0$, Eq.\,\eqref{eq:refined_main} can never be satisfied for any fixed nonzero bias.  Therefore, no domain wall forms.  This is the expected physical result: an infinitely slow biased transition should follow the true vacuum rather than produce a percolating network of metastable domains.  By contrast, the naive estimate \eqref{eq:naive_bound_general} has no such adiabatic limit because it contains no information about the phase transition dynamics.

\paragraph{Comparison between the two criteria.}

Let us make the comparison explicit for the potential \eqref{eq:z2_potential}.  In the small-bias regime, the refined criterion~\eqref{eq:refined_main} becomes
\begin{equation}
 8\delta(T_{\rm fo})\lesssim \frac{L_p}{L_{\rm fo}}.
 \label{eq:refined_delta}
\end{equation}
In terms of the zero-temperature tilt parameter, this is
\begin{equation}
 \frac{8\varepsilon}{\lambda v_0^3}
 \lesssim
 x_{\rm fo}^3\left(\frac{L_p}{L_{\rm fo}}\right),
 \qquad
 x_{\rm fo}\equiv\frac{v(T_{\rm fo})}{v_0}=\sqrt{1-\frac{T_\fo^2}{T_c^2}} \le 1.
 \label{eq:refined_zeroT_delta}
\end{equation}
Even if $x_{\rm fo}=1$, the refined criterion is stronger by a factor of $L_{\rm fo}$. If freeze-out occurs while $v(T_{\rm fo})$ is still below its zero-temperature value, the additional factor $x_{\rm fo}^3$ makes the bound even stronger. 

The freeze-out temperature itself is determined by Eq.\,\eqref{eq:fo-cond}.  Using $\xi(T)\sim 1/[\sqrt{2\lambda}\,v(T)]$ and neglecting the small bias in $V_{b,\fv}$, one finds
\begin{equation}
 \frac{V_{b,\fv}(T)\xi^3(T)}{T}
 \simeq
 \frac{1}{8}\frac{v(T)}{\sqrt{2\lambda}\,T}=\frac{1}{8}\frac{v_0\sqrt{1-T^2/T_c^2} }{\sqrt{2\lambda}T}.
 \label{eq:barrier_exponent_quartic}
\end{equation}
Substituting the above and the Friedmann equation~\eqref{eq:Friedman} into Eq.\,\eqref{eq:fo-cond}, we have
\begin{align}
    \frac{v_0}{8\sqrt{2\lambda}}\frac{\sqrt{1-T_\fo^2/T_c^2}}{T_\fo}=\ln\left(\frac{M_{\rm Pl}}{1.66\sqrt{g_{\star,\rho}}T_\fo}\right).
\end{align}
Considering that the right-hand side depends on $T_\fo$ logarithmically, we can solve this equation by neglecting the $T_\fo$-dependence on the right-hand side. Then one can see that $T_\fo\sim {\rm min}\{T_c,T_v/\bar{L}\}$ where $T_v\equiv v_0/(8\sqrt{2\lambda})$ and 
\begin{align}
    \bar L={\rm max}\Bigg\{\ln\left(\frac{T_c}{H(T_c)}\right), \ln\left(\frac{T_v}{H(T_v)}\right) \Bigg\}.
\end{align}
This gives
\begin{align}
    T_\fo=\sqrt{\frac{T_v^2\,T_c^2 }{T_v^2+\bar{L}^2 T_c^2}},\qquad x_\fo=\frac{1}{\sqrt{(T_v/\bar{L}T_c)^2+1}}.
\end{align}

Substituting the obtained $x_\fo$ into Eq.\,\eqref{eq:refined_main} and replacing $L_\fo$ with $\bar{L}$, we finally arrive at 
\begin{align}
\label{eq:new-cond}
    \frac{8\varepsilon}{\lambda v_0^3}
 &\lesssim \underbrace{\left(\frac{1}{\bar{L}}\frac{1}{[(T_v/\bar{L} T_c)^2+1]^{3/2}}\right)}_{\text{dynamical suppression factor}} L_p.
\end{align}
Compared with the conventional criterion, we now have a dynamical suppression factor. Through the refined criterion, we can identify two parametric regimes:
\begin{equation}
 \frac{8\varepsilon}{\lambda v_0^3}\lesssim
 \begin{cases}
 \bar L^2\left(\dfrac{8\sqrt{2\lambda}T_c}{v_0}\right)^3\times L_p,
 & T_c\ll \dfrac{v_0}{8\sqrt{2\lambda}\bar L},\\[8pt]
 \dfrac{1}{\bar L}\times L_p,
 & T_c\gg \dfrac{v_0}{8\sqrt{2\lambda}\bar L}.
 \end{cases}
 \label{eq:two_regimes}
\end{equation}

\section{Implications for scaling-regime gravitational waves}
\label{sec:scaling-regime}

In discussing gravitational-wave signals from domain walls, one is usually interested in the case where the network enters the so-called scaling regime~\cite{Press:1989yh,Hindmarsh:1996xv,Garagounis:2002kt,Oliveira:2004he,Avelino:2005kn,Leite:2011sc}. In this regime, the formation condition~\eqref{eq:refined_main}, or more explicitly Eq.\,\eqref{eq:new-cond} for the approximate $Z_2$ symmetry breaking, is typically satisfied over much of the parameter space relevant for conventional gravitational-wave studies. Nevertheless, as we show below, the dynamical picture developed here still imposes an additional self-consistency condition on gravitational-wave estimates based on the scaling-regime description.

After entering this regime, the annihilation temperature of the domain walls is estimated as~\cite{Saikawa:2017hiv}\footnote{This formula follows from $t_{\rm ann}\sim 1/\Delta V(0)$, where $t_{\rm ann}$ is the cosmic time at annihilation, while Refs.~\cite{Babichev:2025stm,Dankovsky:2025pjg} find a different behaviour $t_{\rm ann}\sim 1/\Delta V(0)^{2/3}$. The alternative behaviour may need further confirmation with higher-resolution
simulations.}
\begin{align}
    T_{\rm ann}=3.41\times 10^{-2}\,{\rm GeV}\, C_{\rm ann}^{-1/2}\A^{-1/2}\left(\frac{g_{\star,\rho}(T_{\rm ann})}{10}\right)^{-1/4} \left(\frac{\sigma}{\rm TeV^3}\right)^{-1/2}\left(\frac{\Delta V(0)}{\rm MeV^4}\right)^{1/2},
\end{align}
where $C_{\rm ann}$ is an $\O(1)$ coefficient and $\A\simeq 0.8\pm 0.1$ for a $Z_2$ symmetric model~\cite{Hiramatsu:2013qaa}.  Here $\sigma$ is the surface tension of the domain wall. For the approximate $Z_2$ symmetry, we have 
\begin{align}
    \sigma=\int_{-\infty}^\infty \d z\, \left(\frac{\d\phi_{\rm kink}(z)}{\d z}\right)^2=\frac{4}{3}\sqrt{\frac{\lambda}{2}}v_0^3,
\end{align}
where $\phi_{\rm kink}(z)=v_0 {\rm tanh}(v_0z\sqrt{\lambda/2})$ is the kink solution. Assuming that the annihilation occurs during radiation domination, the peak frequency redshifted to today is~\cite{Saikawa:2017hiv}
\begin{align}
    f_{\rm peak}=1.1\times 10^{-9}\,{\rm Hz}\left(\frac{g_{\star,\rho}(T_{\rm ann})}{10}\right)^{1/2}\left(\frac{g_{\star,s}(T_{\rm ann})}{10}\right)^{-1/3}\left(\frac{T_{\rm ann}}{10^{-2}\,{\rm GeV}}\right),
\end{align}
where $g_{\star,s}$ denotes the number of entropy degrees of freedom. The amplitude at the peak frequency is~\cite{Saikawa:2017hiv} 
\begin{align}
    \Omega_{\rm GW}h^2|_{\rm peak}=7.2\times10^{-18} \tilde{\epsilon}\, \A^2 \left(\frac{g_{\star,s}(T_{\rm ann})}{10}\right)^{-4/3}\left(\frac{\sigma}{1\,\rm TeV^3}\right)^2\left(\frac{T_{\rm ann}}{10^{-2}\,\rm GeV}\right)^{-4}\,,
\end{align}
where $\tilde{\epsilon}\simeq 0.7\pm 0.4$ is extracted from numerical simulations~\cite{Hiramatsu:2013qaa}.\footnote{References~\cite{Kitajima:2023kzu,Dankovsky:2025pjg} obtain smaller values of $\tilde{\epsilon}$, around $0.2$. } 

The above formulae are obtained by assuming classical evolution of the domain-wall network, either through analytic estimates or numerical simulations. Therefore, a necessary consistency condition is that the false-vacuum correlation cells freeze out before the walls annihilate, $T_{\fo}>T_{\rm ann}$. Without the refined picture developed in this work, one might naively identify $T_{\fo}$ with $T_c$, in which case this condition could always be satisfied by choosing a sufficiently large $T_c$.

\begin{figure}
    \centering
    \includegraphics[width=0.48\linewidth]{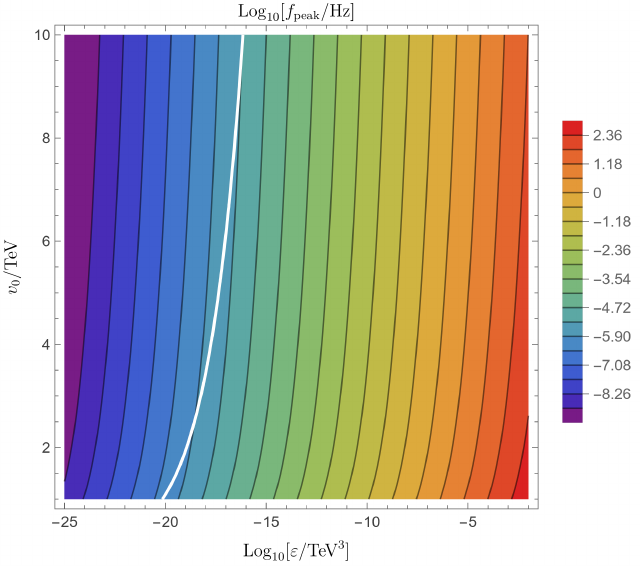}{}\includegraphics[width=0.48\linewidth]{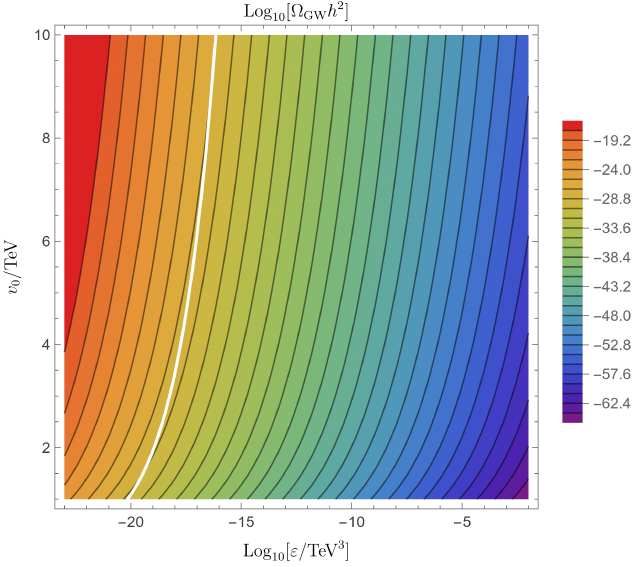}
    \caption{Illustration of the self-consistency condition~\eqref{eq:scaling-regime-cond} for the scaling-regime description. The regions to the right of the white lines violate this condition. The left panel shows the would-be peak frequency, and the right panel shows the would-be peak amplitude, both estimated using the standard scaling-regime formulae. In the regions violating the condition~\eqref{eq:scaling-regime-cond}, these formulae are not expected to give reliable predictions.}
    \label{fig:constraint}
\end{figure}

However, we have shown that $T_{\fo}\sim \min\left\{T_c,T_v/\bar L\right\}$.
Thus, independently of how large $T_c$ is chosen, the scaling-regime description requires the necessary condition
\begin{align}
\label{eq:scaling-regime-cond}
    \frac{v_0}{8\sqrt{2\lambda}\,\bar L}>T_{\rm ann},
    \qquad \text{necessary condition for the scaling regime.}
\end{align}
We illustrate this condition in Fig.\,\ref{fig:constraint}, taking $\lambda=0.5$, $v_0\in[1\,\mathrm{TeV},10\,\mathrm{TeV}]$, $g_{\star,\rho}(T_{\fo})=g_{\star,\rho}(T_{\rm ann})=g_{\star,s}(T_{\rm ann})=100$, $C_{\rm ann}=1$, and $\bar L=40$. The range of $\varepsilon$ is chosen to satisfy both the domination bound and the BBN bound~\cite{Saikawa:2017hiv}. The regions to the right of the white lines violate Eq.\,\eqref{eq:scaling-regime-cond}. In these regions, the domain-wall network cannot enter the scaling regime before annihilation regardless of the value of $T_c$, and the standard scaling-regime estimates for the gravitational-wave signal should not be interpreted as reliable predictions. The actual signal is expected to be further suppressed and is sensitive to the non-scaling dynamics.

\section{Conclusions}
\label{sec:conclusions}

Domain walls provide a well-motivated cosmological source of rich phenomenology, including stochastic gravitational waves, but a biased potential raises a basic question that precedes the subsequent network evolution: whether a percolating domain-wall network forms at all. In this paper, we revisited this question and showed that the criterion conventionally used in the literature misses an essential dynamical effect.

The key point is that false-vacuum correlation cells are not fixed immediately after the false minimum appears. As long as thermal transitions over the barrier are faster than the evolution rate of the background, their population continues to track the instantaneous equilibrium value. The relevant false-vacuum fraction should therefore be evaluated at the freeze-out temperature $T_{\rm fo}$, determined by Eq.\,\eqref{eq:fo-cond}, rather than at the Ginzburg temperature. This leads to the formation criterion~\eqref{eq:refined_main},
which is parametrically stronger than the conventional condition. We have formulated this condition explicitly in terms of model parameters for biased $Z_2$ symmetry breaking. 

The refined criterion has the correct adiabatic behavior: in the limit of an infinitely slow transition, any fixed nonzero bias prevents the formation of a percolating domain-wall network. This shows that biased domain-wall scenarios can be more constrained than previously estimated. In applications to gravitational-wave phenomenology and other cosmological signatures, one should therefore require not only that the walls decay safely after formation, but also that the biased transition satisfies the dynamical formation condition derived in this work. For gravitational-wave phenomenology, the standard scaling-regime estimates further require that annihilation occurs after freeze-out, $T_{\rm ann}<T_{\fo}$. This gives an additional self-consistency condition on the parameter space. If this condition is violated, the walls, if formed, annihilate before entering the scaling regime, and the resulting gravitational-wave signal is controlled by non-scaling dynamics rather than by the standard scaling-regime formulae. Further numerical studies would be useful for testing this criterion beyond the simple mean-field treatment and for extending it to more general potentials and vacuum structures.

\bibliographystyle{utphys}
\bibliography{Refs}

@article{Atta:2026pvw,
    author = "Atta, Utsav and Ghosh, Tathagata and Manna, Sudip",
    title = "{A New Route to the Annihilation of Multi-Wall String Topological Configurations}",
    eprint = "2606.08142",
    archivePrefix = "arXiv",
    primaryClass = "hep-ph",
    reportNumber = "HRI-RECAPP-2026-06",
    month = "6",
    year = "2026"
}

@article{Ma:2025bjf,
    author = "Ma, Ernest and Paul, Partha Kumar and Sahu, Narendra",
    title = "{Lepton parity dark matter and naturally unstable domain walls}",
    eprint = "2508.02642",
    archivePrefix = "arXiv",
    primaryClass = "hep-ph",
    doi = "10.1103/tj6t-dyqn",
    journal = "Phys. Rev. D",
    volume = "112",
    number = "9",
    pages = "095020",
    year = "2025"
}

@article{Borah:2026kfo,
    author = "Borah, Debasish and Paul, Partha Kumar and Sahu, Narendra",
    title = "{Can Dirac neutrinos destabilize $\mathcal{Z}_2$ domain wall network?}",
    eprint = "2602.07380",
    archivePrefix = "arXiv",
    primaryClass = "hep-ph",
    month = "2",
    year = "2026"
}

@article{Datta:2026fav,
    author = "Datta, Satyabrata and Samanta, Rome",
    title = "{PTA-Compatible Domain Walls at LISA and Taiji: Bayesian Reconstruction and Multiband Inference}",
    eprint = "2606.09713",
    archivePrefix = "arXiv",
    primaryClass = "hep-ph",
    month = "6",
    year = "2026"
}

@article{Datta:2026ffs,
    author = "Datta, Satyabrata and Samanta, Rome",
    title = "{Mapping Domain-Wall Bayesian Reconstruction with LISA}",
    eprint = "2606.15387",
    archivePrefix = "arXiv",
    primaryClass = "astro-ph.HE",
    month = "6",
    year = "2026"
}

@article{Masubuchi:2026eau,
    author = "Masubuchi, Haruto and Narita, Yuma and Yin, Wen",
    title = "{Hubble-scale tachyonic shocks from low-scale inflation {\textemdash} A new gravitational-wave window on inflation}",
    eprint = "2602.15825",
    archivePrefix = "arXiv",
    primaryClass = "hep-ph",
    doi = "10.1088/1475-7516/2026/06/090",
    journal = "JCAP",
    volume = "06",
    pages = "090",
    year = "2026"
}

@article{Vanvlasselaer:2026fay,
    author = "Vanvlasselaer, Miguel and Yin, Wen",
    title = "{Spontaneous Baryogenesis from Axions on Induced Electroweak Walls}",
    eprint = "2604.20762",
    archivePrefix = "arXiv",
    primaryClass = "hep-ph",
    month = "4",
    year = "2026"
}

@article{Dankovsky:2025pjg,
    author = "Dankovsky, I. and Ramazanov, S. and Babichev, E. and Gorbunov, D. and Vikman, A.",
    title = "{Cosmic domain walls on a lattice: Illusive effects of initial conditions}",
    eprint = "2509.25367",
    archivePrefix = "arXiv",
    primaryClass = "hep-ph",
    doi = "10.1103/l9gp-7vp5",
    journal = "Phys. Rev. D",
    volume = "112",
    number = "12",
    pages = "123521",
    year = "2025"
}

@article{Kitajima:2023kzu,
    author = "Kitajima, Naoya and Lee, Junseok and Takahashi, Fuminobu and Yin, Wen",
    title = "{Stability of domain walls with inflationary fluctuations under potential bias, and gravitational wave signatures}",
    eprint = "2311.14590",
    archivePrefix = "arXiv",
    primaryClass = "hep-ph",
    reportNumber = "TU-1215",
    doi = "10.1088/1475-7516/2025/07/053",
    journal = "JCAP",
    volume = "07",
    pages = "053",
    year = "2025"
}

@article{Babichev:2025stm,
    author = "Babichev, E. and Dankovsky, I. and Gorbunov, D. and Ramazanov, S. and Vikman, A.",
    title = "{Biased domain walls: faster annihilation, weaker gravitational waves}",
    eprint = "2504.07902",
    archivePrefix = "arXiv",
    primaryClass = "hep-ph",
    doi = "10.1088/1475-7516/2025/10/103",
    journal = "JCAP",
    volume = "10",
    pages = "103",
    year = "2025"
}

@article{Kanemura:2025ixp,
    author = "Kanemura, Shinya and Li, Shao-Ping and Xie, Ke-Pan",
    title = "{Asteroid-mass soliton as the dark matter-baryon coincidence solution}",
    eprint = "2504.08304",
    archivePrefix = "arXiv",
    primaryClass = "hep-ph",
    doi = "10.1103/6tfj-zcls",
    journal = "Phys. Rev. D",
    volume = "112",
    number = "9",
    pages = "L091701",
    year = "2025"
}

@article{Guo:2026cuv,
    author = "Guo, Jinhui and Liu, Jia and Tanaka, Masanori and Wang, Xiao-Ping and Xiao, Huangyu",
    title = "{Supermassive Primordial Black Holes from a Catalyzed Dark Phase Transition for Little Red Dots}",
    eprint = "2604.01304",
    archivePrefix = "arXiv",
    primaryClass = "hep-ph",
    month = "4",
    year = "2026"
}

@article{Sassi:2024cyb,
    author = "Sassi, Mohamed Younes and Moortgat-Pick, Gudrid",
    title = "{Electroweak symmetry restoration in the N2HDM via domain walls}",
    eprint = "2407.14468",
    archivePrefix = "arXiv",
    primaryClass = "hep-ph",
    reportNumber = "DESY-24-109",
    doi = "10.1007/JHEP06(2025)072",
    journal = "JHEP",
    volume = "06",
    pages = "072",
    year = "2025"
}

@article{Hiramatsu:2010yz,
    author = "Hiramatsu, Takashi and Kawasaki, Masahiro and Saikawa, Ken'ichi",
    title = "{Gravitational Waves from Collapsing Domain Walls}",
    eprint = "1002.1555",
    archivePrefix = "arXiv",
    primaryClass = "astro-ph.CO",
    reportNumber = "ICRR-REPORT-559-2009-21, IPMU10-0024",
    doi = "10.1088/1475-7516/2010/05/032",
    journal = "JCAP",
    volume = "05",
    pages = "032",
    year = "2010"
}

@article{Hiramatsu:2013qaa,
    author = "Hiramatsu, Takashi and Kawasaki, Masahiro and Saikawa, Ken'ichi",
    title = "{On the estimation of gravitational wave spectrum from cosmic domain walls}",
    eprint = "1309.5001",
    archivePrefix = "arXiv",
    primaryClass = "astro-ph.CO",
    reportNumber = "ICRR-REPORT-659-2013-8, IPMU13-0182, YITP-13-87",
    doi = "10.1088/1475-7516/2014/02/031",
    journal = "JCAP",
    volume = "02",
    pages = "031",
    year = "2014"
}

@article{Saikawa:2017hiv,
    author = "Saikawa, Ken'ichi",
    title = "{A review of gravitational waves from cosmic domain walls}",
    eprint = "1703.02576",
    archivePrefix = "arXiv",
    primaryClass = "hep-ph",
    reportNumber = "DESY-17-036",
    doi = "10.3390/universe3020040",
    journal = "Universe",
    volume = "3",
    number = "2",
    pages = "40",
    year = "2017"
}

@article{An:2023idh,
    author = "An, Haipeng and Yang, Chen",
    title = "{Gravitational waves produced by domain walls during inflation}",
    eprint = "2304.02361",
    archivePrefix = "arXiv",
    primaryClass = "hep-ph",
    doi = "10.1103/PhysRevD.109.123508",
    journal = "Phys. Rev. D",
    volume = "109",
    number = "12",
    pages = "123508",
    year = "2024"
}

@article{Lu:2024ngi,
    author = "Lu, Bo-Qiang and Chiang, Cheng-Wei and Li, Tianjun",
    title = "{Probing primordial black hole formation from domain wall isocurvature perturbations: constraints and implications}",
    eprint = "2409.09986",
    archivePrefix = "arXiv",
    primaryClass = "astro-ph.CO",
    doi = "10.1088/1475-7516/2026/02/006",
    journal = "JCAP",
    volume = "02",
    pages = "006",
    year = "2026"
}

@article{Dunsky:2024zdo,
    author = "Dunsky, David I. and Kongsore, Marius",
    title = "{Primordial black holes from axion domain wall collapse}",
    eprint = "2402.03426",
    archivePrefix = "arXiv",
    primaryClass = "hep-ph",
    doi = "10.1007/JHEP06(2024)198",
    journal = "JHEP",
    volume = "06",
    pages = "198",
    year = "2024"
}

@article{Lee:2025yvn,
    author = "Lee, Junseok and Murai, Kai and Takahashi, Fuminobu and Yin, Wen",
    title = "{Isotropic cosmic birefringence from string axion domain walls without cosmic strings and DESI results}",
    eprint = "2503.18417",
    archivePrefix = "arXiv",
    primaryClass = "hep-ph",
    reportNumber = "TU-1259",
    doi = "10.1103/pjf4-sn4v",
    journal = "Phys. Rev. D",
    volume = "112",
    number = "4",
    pages = "043538",
    year = "2025"
}

@article{Roshan:2026yon,
    author = "Roshan, Rishav",
    title = "{Imprint of domain wall annihilation on induced gravitational waves}",
    eprint = "2604.25726",
    archivePrefix = "arXiv",
    primaryClass = "hep-ph",
    month = "4",
    year = "2026"
}

@article{Vilenkin:1984ib,
    author = "Vilenkin, Alexander",
    title = "{Cosmic Strings and Domain Walls}",
    reportNumber = "PRINT-84-0840 (TUFTS)",
    doi = "10.1016/0370-1573(85)90033-X",
    journal = "Phys. Rept.",
    volume = "121",
    pages = "263--315",
    year = "1985"
}

@article{Liu:2019lul,
    author = "Liu, Jing and Guo, Zong-Kuan and Cai, Rong-Gen",
    title = "{Primordial Black Holes from Cosmic Domain Walls}",
    eprint = "1908.02662",
    archivePrefix = "arXiv",
    primaryClass = "astro-ph.CO",
    doi = "10.1103/PhysRevD.101.023513",
    journal = "Phys. Rev. D",
    volume = "101",
    number = "2",
    pages = "023513",
    year = "2020"
}

@article{Li:2025gld,
    author = "Li, Yuan-Jie and Liu, Jing and Guo, Zong-Kuan",
    title = "{Dynamics of $Z_N$ domain walls with bias directions}",
    eprint = "2502.13644",
    archivePrefix = "arXiv",
    primaryClass = "astro-ph.CO",
    doi = "10.1103/rnpp-7wh2",
    journal = "Phys. Rev. D",
    volume = "112",
    number = "10",
    pages = "103510",
    year = "2025"
}

@article{Wei:2022poh,
    author = "Wei, Dongdong and Jiang, Yun",
    title = "{Domain wall networks from first-order phase transitions and gravitational waves}",
    eprint = "2208.07186",
    archivePrefix = "arXiv",
    primaryClass = "hep-ph",
    doi = "10.1103/PhysRevD.110.123505",
    journal = "Phys. Rev. D",
    volume = "110",
    number = "12",
    pages = "123505",
    year = "2024"
}

@article{Borah:2025bfa,
    author = "Borah, Debasish and Saha, Indrajit",
    title = "{Gravitational waves from seesaw assisted collapsing domain walls}",
    eprint = "2512.22339",
    archivePrefix = "arXiv",
    primaryClass = "hep-ph",
    month = "12",
    year = "2025"
}

@article{Azzola:2026cwa,
    author = "Azzola, Jacopo and Matsedonskyi, Oleksii and Weiler, Andreas",
    title = "{Baryon Asymmetry from Electroweak-Symmetric Domain Walls}",
    eprint = "2604.16603",
    archivePrefix = "arXiv",
    primaryClass = "hep-ph",
    month = "4",
    year = "2026"
}

@article{Bai:2026udq,
    author = "Bai, Yang and Lyu, Kun-Feng and Zhao, Yue",
    title = "{Electroweak Baryogenesis from Collapsing Domain Walls}",
    eprint = "2604.27376",
    archivePrefix = "arXiv",
    primaryClass = "hep-ph",
    month = "4",
    year = "2026"
}

@article{Blasi:2025tmn,
    author = {Blasi, Simone and Mariotti, Alberto and Rase, A{\"a}ron and Vanvlasselaer, Miguel},
    title = "{Domain walls in the scaling regime: Equal Time Correlator and gravitational waves}",
    eprint = "2511.16649",
    archivePrefix = "arXiv",
    primaryClass = "hep-ph",
    doi = "10.1088/1475-7516/2026/06/053",
    journal = "JCAP",
    volume = "06",
    pages = "053",
    year = "2026"
}

@article{Gouttenoire:2023gbn,
    author = "Gouttenoire, Yann and Vitagliano, Edoardo",
    title = "{Primordial black holes and wormholes from domain wall networks}",
    eprint = "2311.07670",
    archivePrefix = "arXiv",
    primaryClass = "hep-ph",
    doi = "10.1103/PhysRevD.109.123507",
    journal = "Phys. Rev. D",
    volume = "109",
    number = "12",
    pages = "123507",
    year = "2024"
}

@article{Blasi:2022woz,
    author = "Blasi, Simone and Mariotti, Alberto",
    title = "{Domain Walls Seeding the Electroweak Phase Transition}",
    eprint = "2203.16450",
    archivePrefix = "arXiv",
    primaryClass = "hep-ph",
    doi = "10.1103/PhysRevLett.129.261303",
    journal = "Phys. Rev. Lett.",
    volume = "129",
    number = "26",
    pages = "261303",
    year = "2022"
}

@article{Blasi:2023sej,
    author = {Blasi, Simone and Mariotti, Alberto and Rase, A{\"a}ron and Sevrin, Alexander},
    title = "{Axionic domain walls at Pulsar Timing Arrays: QCD bias and particle friction}",
    eprint = "2306.17830",
    archivePrefix = "arXiv",
    primaryClass = "hep-ph",
    doi = "10.1007/JHEP11(2023)169",
    journal = "JHEP",
    volume = "11",
    pages = "169",
    year = "2023"
}

@article{Gouttenoire:2025ofv,
    author = "Gouttenoire, Yann and King, Stephen F. and Roshan, Rishav and Wang, Xin and White, Graham and Yamazaki, Masahito",
    title = "{Cosmological consequences of domain walls biased by quantum gravity}",
    eprint = "2501.16414",
    archivePrefix = "arXiv",
    primaryClass = "hep-ph",
    doi = "10.1103/7zmx-v16z",
    journal = "Phys. Rev. D",
    volume = "112",
    number = "7",
    pages = "075007",
    year = "2025"
}

@article{Sassi:2023cqp,
    author = "Sassi, Mohamed Younes and Moortgat-Pick, Gudrid",
    title = "{Domain walls in the Two-Higgs-Doublet Model and their charge and CP-violating interactions with Standard Model fermions}",
    eprint = "2309.12398",
    archivePrefix = "arXiv",
    primaryClass = "hep-ph",
    reportNumber = "DESY-23-134",
    doi = "10.1007/JHEP04(2024)101",
    journal = "JHEP",
    volume = "04",
    pages = "101",
    year = "2024"
}

@article{Wu:2022tpe,
    author = "Wu, Yongcheng and Xie, Ke-Pan and Zhou, Ye-Ling",
    title = "{Classification of Abelian domain walls}",
    eprint = "2205.11529",
    archivePrefix = "arXiv",
    primaryClass = "hep-ph",
    doi = "10.1103/PhysRevD.106.075019",
    journal = "Phys. Rev. D",
    volume = "106",
    number = "7",
    pages = "075019",
    year = "2022"
}

@article{Li:2023gil,
    author = "Li, Yang and Bian, Ligong and Cai, Rong-Gen and Shu, Jing",
    title = "{Cosmic simulations of axion: probing dark matter and gravitational waves}",
    eprint = "2311.02011",
    archivePrefix = "arXiv",
    primaryClass = "astro-ph.CO",
    doi = "10.1088/1475-7516/2025/08/091",
    journal = "JCAP",
    volume = "08",
    pages = "091",
    year = "2025"
}

@article{Wu:2022stu,
    author = "Wu, Yongcheng and Xie, Ke-Pan and Zhou, Ye-Ling",
    title = "{Collapsing domain walls beyond Z2}",
    eprint = "2204.04374",
    archivePrefix = "arXiv",
    primaryClass = "hep-ph",
    doi = "10.1103/PhysRevD.105.095013",
    journal = "Phys. Rev. D",
    volume = "105",
    number = "9",
    pages = "095013",
    year = "2022"
}

@article{Fu:2025qhf,
    author = "Fu, Bowen and King, Stephen F. and Marsili, Luca and Turner, Jessica and Zhou, Ye-Ling",
    title = "{Domain Walls in $A_4$ Flavour Models}",
    eprint = "2512.13784",
    archivePrefix = "arXiv",
    primaryClass = "hep-ph",
    reportNumber = "IPPP/25/80",
    month = "12",
    year = "2025"
}

@article{Ghoshal:2025gci,
    author = "Ghoshal, Anish and Hamada, Yu",
    title = "{Gravitational waves created by current-carrying domain walls}",
    eprint = "2501.01542",
    archivePrefix = "arXiv",
    primaryClass = "hep-ph",
    reportNumber = "DESY-24-198",
    doi = "10.1103/24w5-rflt",
    journal = "Phys. Rev. D",
    volume = "112",
    number = "11",
    pages = "115019",
    year = "2025"
}

@article{Krajewski:2021jje,
    author = "Krajewski, Tomasz and Kwapisz, Jan Henryk and Lalak, Zygmunt and Lewicki, Marek",
    title = "{Stability of domain walls in models with asymmetric potentials}",
    eprint = "2103.03225",
    archivePrefix = "arXiv",
    primaryClass = "astro-ph.CO",
    doi = "10.1103/PhysRevD.104.123522",
    journal = "Phys. Rev. D",
    volume = "104",
    number = "12",
    pages = "123522",
    year = "2021"
}

@article{Correia:2018tty,
    author = "Correia, J. R. C. C. C. and Leite, I. S. C. R. and Martins, C. J. A. P.",
    title = "{Effects of biases in domain wall network evolution. II. Quantitative analysis}",
    eprint = "1804.10761",
    archivePrefix = "arXiv",
    primaryClass = "astro-ph.CO",
    doi = "10.1103/PhysRevD.97.083521",
    journal = "Phys. Rev. D",
    volume = "97",
    number = "8",
    pages = "083521",
    year = "2018"
}

@book{vilenkin1994cosmic,
  title={Cosmic strings and other topological defects},
  author={Alexander and Shellard, EPS},
  year={1994},
  publisher={Cambridge University Press}
}

@article{Fujikura:2024biv,
    author = "Fujikura, Kohei and Sakellariadou, Mairi and Uwabo-Niibo, Michiru and Yamaguchi, Masahide",
    title = "{Formation of defects associated with both spontaneous and explicit symmetry breaking}",
    eprint = "2410.07565",
    archivePrefix = "arXiv",
    primaryClass = "hep-ph",
    reportNumber = "KCL-PH-TH-2024-51, UT-Komaba/23-11",
    doi = "10.1103/PhysRevD.111.023511",
    journal = "Phys. Rev. D",
    volume = "111",
    number = "2",
    pages = "023511",
    year = "2025"
}

@article{Hiramatsu:2012sc,
    author = "Hiramatsu, Takashi and Kawasaki, Masahiro and Saikawa, Ken'ichi and Sekiguchi, Toyokazu",
    title = "{Axion cosmology with long-lived domain walls}",
    eprint = "1207.3166",
    archivePrefix = "arXiv",
    primaryClass = "hep-ph",
    reportNumber = "ICRR-REPORT-620-2012-9, IPMU12-0140, YITP-12-58",
    doi = "10.1088/1475-7516/2013/01/001",
    journal = "JCAP",
    volume = "01",
    pages = "001",
    year = "2013"
}

@article{Correia:2014kqa,
    author = "Correia, J. R. C. C. C. and Leite, I. S. C. R. and Martins, C. J. A. P.",
    title = "{Effects of Biases in Domain Wall Network Evolution}",
    eprint = "1407.3905",
    archivePrefix = "arXiv",
    primaryClass = "hep-ph",
    doi = "10.1103/PhysRevD.90.023521",
    journal = "Phys. Rev. D",
    volume = "90",
    number = "2",
    pages = "023521",
    year = "2014"
}

@article{Coulson:1995nv,
    author = "Coulson, D. and Lalak, Z. and Ovrut, Burt A.",
    title = "{Biased domain walls}",
    reportNumber = "UPR-0668-T",
    doi = "10.1103/PhysRevD.53.4237",
    journal = "Phys. Rev. D",
    volume = "53",
    pages = "4237--4246",
    year = "1996"
}

@article{Vilenkin:1982ks,
    author = "Vilenkin, A. and Everett, A. E.",
    title = "{Cosmic Strings and Domain Walls in Models with Goldstone and PseudoGoldstone Bosons}",
    doi = "10.1103/PhysRevLett.48.1867",
    journal = "Phys. Rev. Lett.",
    volume = "48",
    pages = "1867--1870",
    year = "1982"
}

@article{Sikivie:1982qv,
    author = "Sikivie, P.",
    title = "{Of Axions, Domain Walls and the Early Universe}",
    reportNumber = "UFTP-82-3",
    doi = "10.1103/PhysRevLett.48.1156",
    journal = "Phys. Rev. Lett.",
    volume = "48",
    pages = "1156--1159",
    year = "1982"
}

@article{Vilenkin:1981zs,
    author = "Vilenkin, A.",
    title = "{Gravitational Field of Vacuum Domain Walls and Strings}",
    doi = "10.1103/PhysRevD.23.852",
    journal = "Phys. Rev. D",
    volume = "23",
    pages = "852--857",
    year = "1981"
}

@article{Rams:2019njq,
    author = "Rams, Marek M. and Dziarmaga, Jacek and Zurek, Wojciech H.",
    title = "{Symmetry Breaking Bias and the Dynamics of a Quantum Phase Transition}",
    eprint = "1905.05783",
    archivePrefix = "arXiv",
    primaryClass = "cond-mat.stat-mech",
    doi = "10.1103/PhysRevLett.123.130603",
    journal = "Phys. Rev. Lett.",
    volume = "123",
    number = "13",
    pages = "130603",
    year = "2019"
}

@article{Rysti:2019fwr,
    author = {Rysti, J. and M{\"a}kinen, J. T. and Autti, S. and Kamppinen, T. and Volovik, G. E. and Eltsov, V. B.},
    title = "{Suppressing the Kibble-Zurek Mechanism by a Symmetry-Violating Bias}",
    eprint = "1906.11453",
    archivePrefix = "arXiv",
    primaryClass = "cond-mat.other",
    doi = "10.1103/PhysRevLett.127.115702",
    journal = "Phys. Rev. Lett.",
    volume = "127",
    number = "11",
    pages = "115702",
    year = "2021"
}

@article{Larsson:1996sp,
    author = "Larsson, Sebastian E. and Sarkar, Subir and White, Peter L.",
    title = "{Evading the cosmological domain wall problem}",
    eprint = "hep-ph/9608319",
    archivePrefix = "arXiv",
    reportNumber = "OUTP-96-11-P",
    doi = "10.1103/PhysRevD.55.5129",
    journal = "Phys. Rev. D",
    volume = "55",
    pages = "5129--5135",
    year = "1997"
}

@article{Gelmini:1988sf,
    author = "Gelmini, Graciela B. and Gleiser, Marcelo and Kolb, Edward W.",
    title = "{Cosmology of Biased Discrete Symmetry Breaking}",
    reportNumber = "NSF-ITP-88-148, FERMILAB-PUB-88-151-A",
    doi = "10.1103/PhysRevD.39.1558",
    journal = "Phys. Rev. D",
    volume = "39",
    pages = "1558",
    year = "1989"
}

@article{Leite:2011sc,
    author = "Leite, A. M. M. and Martins, C. J. A. P.",
    title = "{Scaling Properties of Domain Wall Networks}",
    eprint = "1110.3486",
    archivePrefix = "arXiv",
    primaryClass = "hep-ph",
    doi = "10.1103/PhysRevD.84.103523",
    journal = "Phys. Rev. D",
    volume = "84",
    pages = "103523",
    year = "2011"
}

@article{Hindmarsh:1996xv,
    author = "Hindmarsh, Mark",
    title = "{Analytic scaling solutions for cosmic domain walls}",
    eprint = "hep-ph/9605332",
    archivePrefix = "arXiv",
    reportNumber = "SUSX-TH-96-005",
    doi = "10.1103/PhysRevLett.77.4495",
    journal = "Phys. Rev. Lett.",
    volume = "77",
    pages = "4495--4498",
    year = "1996"
}

@article{Avelino:2005kn,
    author = "Avelino, P. P. and Martins, C. J. A. P. and Oliveira, J. C. R. E.",
    title = "{One-scale model for domain wall network evolution}",
    eprint = "hep-ph/0507272",
    archivePrefix = "arXiv",
    doi = "10.1103/PhysRevD.72.083506",
    journal = "Phys. Rev. D",
    volume = "72",
    pages = "083506",
    year = "2005"
}

@article{Oliveira:2004he,
    author = "Oliveira, J. C. R. E. and Martins, C. J. A. P. and Avelino, P. P.",
    title = "{The Cosmological evolution of domain wall networks}",
    eprint = "hep-ph/0410356",
    archivePrefix = "arXiv",
    doi = "10.1103/PhysRevD.71.083509",
    journal = "Phys. Rev. D",
    volume = "71",
    pages = "083509",
    year = "2005"
}

@article{Garagounis:2002kt,
    author = "Garagounis, Theodore and Hindmarsh, Mark",
    title = "{Scaling in numerical simulations of domain walls}",
    eprint = "hep-ph/0212359",
    archivePrefix = "arXiv",
    reportNumber = "SUSX-TH-02-029",
    doi = "10.1103/PhysRevD.68.103506",
    journal = "Phys. Rev. D",
    volume = "68",
    pages = "103506",
    year = "2003"
}

@article{Press:1989yh,
    author = "Press, William H. and Ryden, Barbara S. and Spergel, David N.",
    title = "{Dynamical Evolution of Domain Walls in an Expanding Universe}",
    reportNumber = "NSF-ITP-89-51, CFA-1870",
    doi = "10.1086/168151",
    journal = "Astrophys. J.",
    volume = "347",
    pages = "590--604",
    year = "1989"
}

@article{Kibble:1976sj,
    author = "Kibble, T. W. B.",
    title = "{Topology of Cosmic Domains and Strings}",
    reportNumber = "ICTP/75/5",
    doi = "10.1088/0305-4470/9/8/029",
    journal = "J. Phys. A",
    volume = "9",
    pages = "1387--1398",
    year = "1976"
}

@article{Zeldovich:1974uw,
    author = "Zeldovich, Ya. B. and Kobzarev, I. Yu. and Okun, L. B.",
    title = "{Cosmological Consequences of the Spontaneous Breakdown of Discrete Symmetry}",
    reportNumber = "SLAC-TRANS-0165, IPM-MOSCOW-15",
    journal = "Zh. Eksp. Teor. Fiz.",
    volume = "67",
    pages = "3--11",
    year = "1974"
}

\end{document}